\documentclass{article}
\usepackage{spconf,amsmath,graphicx, url}
\usepackage{hyperref}
\usepackage[T1]{fontenc}
\usepackage{multirow}
\usepackage{amssymb}
\usepackage{color}
\usepackage{algorithm}
\usepackage{algpseudocode}
\usepackage{cite}
\usepackage{multirow}
\usepackage[table,xcdraw]{xcolor}
\usepackage{threeparttable}
\newcommand{\rongkai}[1]{\textcolor{black}{#1}}
\newcommand{\bihan}[1]{\textcolor{black}{#1}}

\title{RePnP: Plug-and-Play with Deep Reinforcement Learning Prior for Robust Image restoration}
\name{\vspace{-0.1in} Chong Wang$^{1*}$,~Rongkai Zhang$^{1*}$,~Saiprasad Ravishankar$^2$ and Bihan Wen$^{1\dagger}$ \thanks{$^*$Both authors contributed equally to this research.}\thanks{$^\dagger$Bihan Wen is the corresponding author. This research is supported by Academic Research Fund Tier 1 (Project ID: RG137/20), MOE, Singapore.}}
\address{$^1$School of Electrical \& Electronic Engineering, Nanyang Technological University, Singapore.\\
$^2$Departments of Computational Mathematics, Science and Engineering, \\
  and Biomedical Engineering, Michigan State University, East Lansing, MI, USA.}
\begin{document}
\ninept
\maketitle
\begin{abstract}
Image restoration schemes based on the pre-trained deep models have received great attention due to their unique flexibility for solving various inverse problems.
In particular, the Plug-and-Play (PnP) framework is a popular and powerful tool that can integrate an off-the-shelf deep denoiser for different image restoration tasks with known observation models.
However, obtaining the observation model that exactly matches the actual one can be challenging in practice. Thus, the PnP schemes with conventional deep denoisers may fail to generate satisfying results in some real-world image restoration tasks.
We argue that the robustness of the PnP framework is largely limited by using the off-the-shelf deep denoisers that are trained by deterministic optimization.
To this end, we propose a novel deep reinforcement learning (DRL) based PnP framework, dubbed RePNP, by leveraging a light-weight DRL-based denoiser for robust image restoration tasks.
Experimental results demonstrate that the proposed RePNP is robust to 
the observation model used in the PnP scheme deviating from the actual one.
Thus, RePNP can generate more reliable restoration results for image deblurring and super resolution tasks. 
Compared with several state-of-the-art deep image restoration baselines, RePNP achieves better results subjective to model deviation with fewer model parameters.
\end{abstract}
%

\section{Introduction} \label{sec1}
Image restoration (IR) \bihan{aims to recover}
the latent clean image $x$ from its degraded measurements $y$ and these are typically related by the following model:
\begin{equation}\label{eq1}
    y = Hx + n,
\end{equation}
where $H$ and $n$ denote the forward model and additive noise, respectively.
\bihan{Different IR tasks are associated with corresponding observation models $H$, e.g., in image deblurring and super-resolution tasks, the $H$ is a blurring operator and decimation operator (blurring followed by downsampling), respectively.}
\bihan{As 
IR using~\eqref{eq1} is typically
ill-posed, different prior-based methods have been proposed to improve image recovery, including non-local algorithms \cite{BM3D}, sparse coding schemes~\cite{ksvd, wen2020image} and recently deep learning-based methods \cite{dncnn, rdn, liu2018non}}

\bihan{Recently, the model-based deep learning methods, by incorporating the known $H$ into the deep networks, have demonstrated remarkable performance on IR and imaging tasks.}
\bihan{For example, the LISTA scheme~\cite{lista} proposed to unfold the iterative shrinkage thresholding algorithm (ISTA)~\cite{ista} for sparse coding, which demonstrated superior results in IR tasks such as super-resolution~\cite{liu2016robust}.}
\bihan{More recently, the Plug-and-Play (PnP) scheme~\cite{pnp} has gained great popularity due to its effectiveness and flexibility for solving a wide range of inverse problems, e.g., medical imaging~\cite{pnpct}, snapshot compressive imaging~\cite{pnptvsci, pnpsci} and image restoration \cite{pnpfp, ircnn, dpir}.
Different from many unrolling methods, PnP directly adopts an off-the-shelf deep denoiser to replace the proximal mapping (involving the regularizer) inside an iterative reconstruction algorithm. Thus, one can easily apply PnP algorithms for future IR tasks given $H$, without additional data annotation or training.}
\bihan{However, the PnP schemes commonly require accurate choice of $H$~\cite{pnpfp, dpir}, which cannot always be guaranteed in practice.
Any bias or deviation of the chosen $H$ (from the true model) may lead to degraded reconstruction results using the conventional PnP image reconstruction scheme.} 

\bihan{In this work, we investigate and improve the robustness of PnP under a more practical yet challenging setting, i.e., the $H$ used for PnP deviates from the actual one. We show that the PnP schemes, by incorporating a deep denoiser that is trained by supervised learning with deterministic optimization, fail to generate satisfying results for IR tasks.}
\rongkai{To this end, we propose a novel image restoration scheme by incorporating a deep reinforcement learning (DRL) denoiser into the PnP framework, dubbed RePNP.}
\bihan{Different from the conventional deep denoisers, the DRL-based denoiser models the image recovery task as a Markov decision process.}
\bihan{Besides, the proposed DRL denoiser has much fewer network parameters compared to other popular or state-of-the-art supervised deep image denoisers.}
\bihan{To the best of our knowledge, no work to date has studied and integrated a DRL denoiser with the PnP scheme.}
\bihan{We conduct extensive experiments and demonstrate that the proposed RePNP approach can provide more robust restoration results, with respect to the deviation of $H$, in image deblurring and super-resolution tasks, compared to the conventional PnP schemes using state-of-the-art deep denoisers.}


\section{Related Work} \label{section2}

\subsection{Plug-and-Play for Image Restoration}
Recently, deep learning based methods have witnessed a dramatic increase in popularity for inverse problem tasks~\cite{dncnn,liu2018non}.
\bihan{However, most of the conventional deep models are trained for one specific IR task, and require re-training when extending to other tasks.}
By splitting the variables involved in the forward model and the regularizer, 
the Plug-and-Play (PnP) methods have attracted significant attention due to their unique flexibility for solving different inverse problems~\cite{pnp}.
\bihan{PnP utilizes various denoisers, such as K-SVD~\cite{ksvd} or BM3D~\cite{BM3D} as a replacement for the proximal operator corresponding to the regularizer in ADMM~\cite{admm} iterations.
}
Apart from ADMM, other variable splitting algorithms, such as half quadratic splitting~\cite{hqs}, have been adopted in the PnP paradigms~\cite{pnphqs, dpir}.
More recent PnP methods achieved promising IR results with the help of state-of-the-art deep image denoisers~\cite{ircnn, pnpsci, pnptvsci}.
For example, Qiu et al.~\cite{ffdtvpnp} combined a classical denoiser (TV~\cite{tv}) and a deep denoiser (FFDNet~\cite{ffdnet}) in PnP to boost the quality of reconstructed images.
Besides, several theoretical analyses on convergence of PnP methods to a fixed point have been investigated~\cite{pnpfp}. 

\vspace{-0.1in}
\subsection{DRL for Image Denoising}
Apart from the primary application of DRL in many domains, it has been extended to some computer vision tasks in recent years.
Zhang et al.~\cite{rellie} model low-light image enhancement as a markov decision process and learn a policy that provides customized enhanced outputs by flexibly applying the policy multiple times.
Yu et al.~\cite{rlr} learn a policy to select an appropriate CNN from a pre-defined toolbox to progressively restore high-quality images.
Instead of operating on the whole image, Furuta et al.~\cite{pixrl} achieve the pixel-wise denoising via assigning every pixel an agent and the action is selected from a set of pre-defined basic filters, e.g., Gaussian filters.
A similar idea was proposed by Zhang et al.~\cite{r3l}, where they replace the action set by a range of pixel value modification steps to directly learn a residual noise map for image denoising.

\section{DRL-based Plug-and-Play Scheme}
\label{section3}

\subsection{Preliminary on Plug-and-Play Algorithms}
\bihan{To solve the ill-posed IR problem~\eqref{eq1}, the widely-used maximum a posteriori (MAP) approach estimates the clean image $x$ using the following formulation:}
\begin{equation}
    (\hat{x},\hat{z} ) = \arg\min_{x, z}{\frac{1}{2}\lVert{Hx-y}\rVert^2_2+\lambda\mathcal{R}{(z)}} \quad \text{s.t.} \;  x=z, \label{eq2}
\end{equation}
where $\mathcal{R}(\cdot)$ is 
\bihan{the regularizer} capturing assumed image properties and 
$z$ is an auxiliary variable that makes \eqref{eq2} a constrained optimization problem.
Using the half quadratic splitting algorithm~\cite{hqs}, the data fidelity term and prior term can be decoupled, converting \eqref{eq2} to the two sub-problems as follows:
\begin{align}
    x^{(k+1)} &= \arg\min_{x}{\lVert{Hx-y}\rVert^2_2+\mu \lVert{x-z^{(k)}}\rVert^2_2}, \label{eq5}\\
    z^{(k+1)} &= \arg\min_{z}\frac{\mu}{2}\lVert{z-x^{(k+1)}}\rVert^2_2 + \lambda\mathcal{R}{(z)}.\label{eq6}
\end{align}
\bihan{Here, (\ref{eq5}) is a least-squares problem with a quadratic penalty term, which has a closed-form solution}
\begin{equation}
    x^{(k+1)} = (H^TH+\mu I)^{-1}(H^Ty+\mu z^{(k)}). \label{eq7}
\end{equation}
The matrix inversion in \eqref{eq7} would be computationally expensive when the forward operator $H$ is large. In practice, one can apply numerical algorithms (e.g., conjugate gradients) for obtaining a solution efficiently~\cite{onlinepnp}. 
Besides, \eqref{eq6} can be viewed as a denoising problem. Thus, instead of explicitly formulating the $\mathcal{R}(\cdot)$, the solution to \eqref{eq6} can be obtained by a denoising step:
\begin{equation}
    z^{(k+1)} = \mathcal{D_\sigma}(x^{(k+1)}), \label{eq8}
\end{equation}
where $\mathcal{D_\sigma}$ denotes a denoiser
with a standard deviation of $\sigma$. 

\subsection{Overview of RePNP Framework}
\begin{figure}[t]
\begin{center}
\includegraphics[height=1.65in]{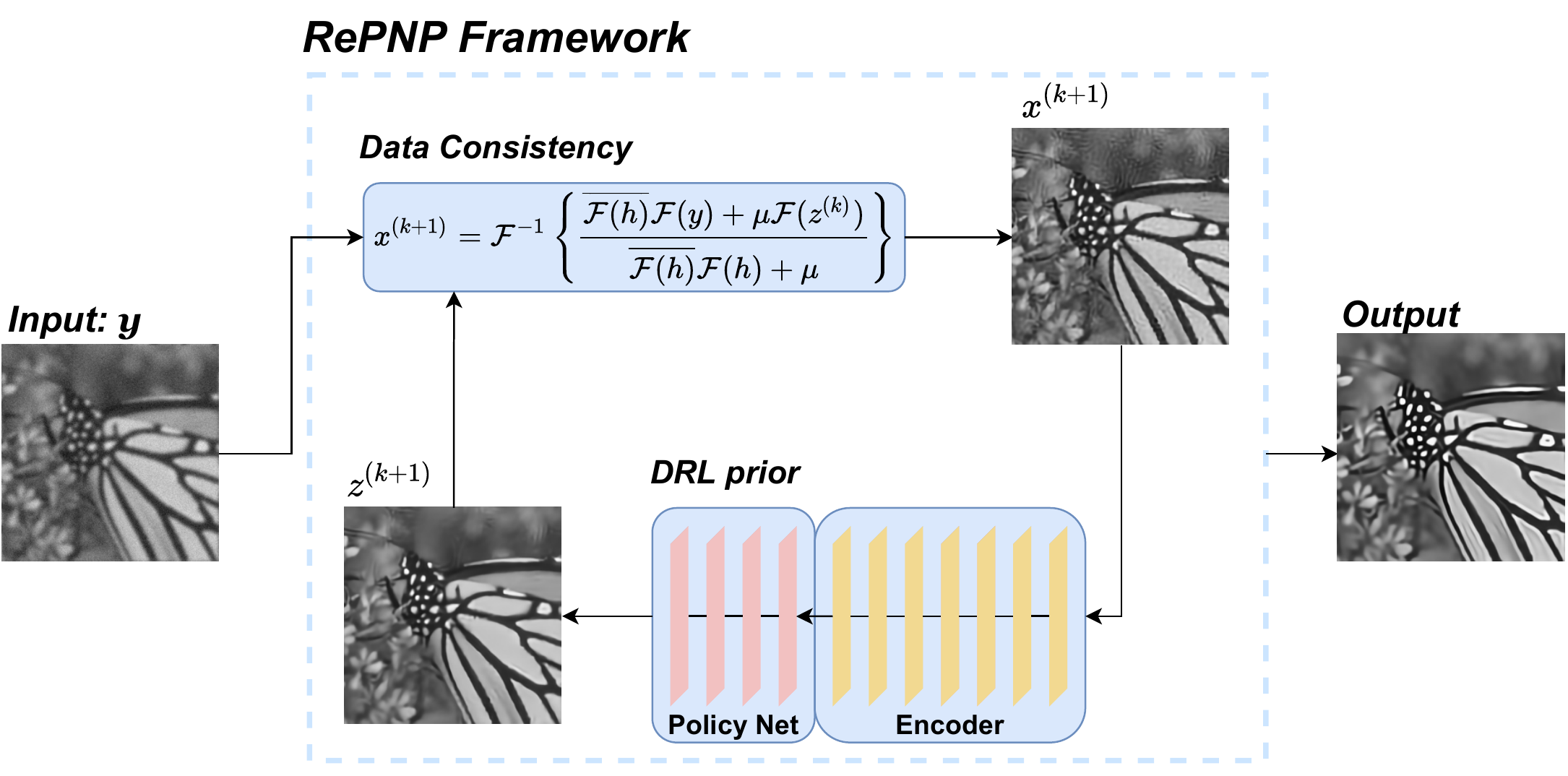}
\caption{The overall architecture of the proposed RePNP framework for image deblurring. Specifically, the data consistency module is a fast implementation of~\eqref{eq7}, where $\mathcal{F}(\cdot)$ and $\overline{\mathcal{F}(\cdot)}$ are the Fourier transform operator and its the complex conjugate respectively and $h$ is the finite impulse response filter representing the blur kernel.}
\label{flow chart}
\end{center}
\vspace{-0.3in}
\end{figure}

Fig.~\ref{flow chart} illustrates the overall architecture of our proposed RePNP algorithm for image deblurring. 
Different from conventional PnP methods, where supervised learning based deep denoisers are adopted, we integrate the PnP framework with a DRL-based denoiser that iteratively removes noise with a sequence of pixel-wise modifications. 
By training the network via a stochastic state-wise reward, a more general policy can be obtained. Being able to explore more possible states, our DRL denoiser based PnP method is more robust when facing various types of noise and artifacts compared to other conventional PnP methods.
During the PnP iteration, our DRL denoiser consists of an encoder network followed by a policy network and is much more lightweight compared with other deep network denoisers. 
\subsection{Image Denoising as Markov Decision Process}
Image denoising can be achieved by sequentially removing the residuals from the noisy images. This process can be naturally modeled as a Markov Decision Process. Particularly, taking the noisy image $I^t\in \mathbb{R}^N$ as the state at time step $t$, the policy $\pi$ will output a probability map $P(a^t_i|I^t)$ for every pixel, where $a^t_i$ denotes the action for the $i$-th pixel, i.e., the estimated residual to be removed at that pixel. The action belongs to a predefined action set $A$, which consists of a range of discrete pixel values. 
By updating $I^t$ with the action map $a^t$, the network can generate a new estimate of the clean image $I^{t+1}$ and it iterates until the termination state $t = T$. 
The probability of an action trajectory $J_i^{t}\triangleq\{a_i^1, a_i^2, \cdots, a_i^t\}$ at pixel index $i$, denoted as $P(J_i^t|I^0,\theta_\pi)$, can be calculated as:
\begin{equation}
    P(J_i^t|I^0,\theta_\pi) = \prod_{t=1}^T{P(a_i^t|J_i^{t-1},I^0,\theta_\pi)}. \label{eq9}
\end{equation}
When obtaining a new denoised estimate $I^{t+1}$, the agent can calculate a reward $r_i^t$ at each pixel $I_i$ as the step-wise improvement:
\begin{equation}
    r_i^t \triangleq (x_i-I_i^{t-1})^2-(x_i-I_i^t)^2,
\end{equation}
where $x_i$ is the clean image pixel intensity. The discounted long-term reward $R_i^t(J_i)$ to evaluate the policy at state $t$ is defined as:
\begin{equation}
    R_i^t(J_i^t)\triangleq r_i^{t} + \gamma r_i^{t+1} + \gamma^2 r_i^{t+2} + \cdots + \gamma^{T-t}r_i^{T}, \label{eq10}
\end{equation}
where $\gamma^j$ denotes the $j$-th power of the discount factor $\gamma \in (0,1]$.

\begin{figure}[!t]
\begin{center}
\includegraphics[height=2.0in]{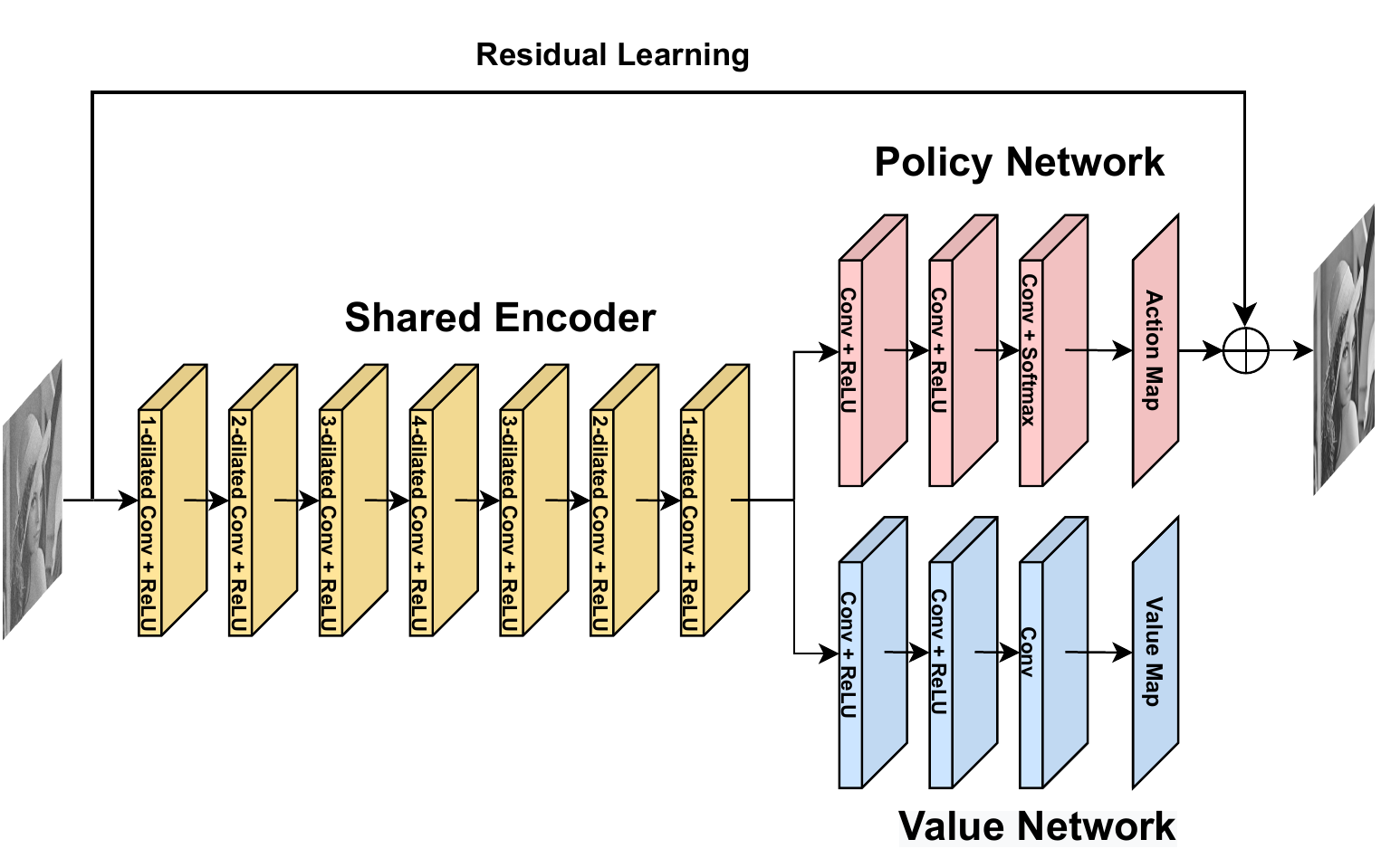}
\caption{Network architecture of the proposed DRL denoiser during the training process.}
\label{network architecture}
\end{center}
\vspace{-0.3in}
\end{figure}

\subsection{Proposed DRL Denoiser}
The architecture of our proposed DRL denoiser is illustrated in Fig.~\ref{network architecture}.
We apply a fully convolutional network (FCN) based proximal policy optimization (PPO)~\cite{PPO} framework as the DRL denoiser. We adopt FCN as the backbone, since it is widely used and effective in image processing tasks for pixel-level modification.
Instead of directly taking $I^t$ as the input to the policy network, we use a shared encoder whose parameters are updated as part of the policy network to get the state representation $s^t$. Then we train a policy network $\pi(\cdot|s^t, \theta_\pi)$ and a value network $V(\cdot|s^t, \theta_V)$ with the help of PPO to make the training more stable and efficient. 
Taking $s^t$, the policy network $\pi(\cdot|s^t, \theta_\pi)$ outputs the probability of selecting a certain action $a^t_i$ from a pre-defined range [-13, 13] with integer gradation, which is designed to cover a wide range of noise while keeping the searching space narrow. The value network outputs $v^t_i$, which is the estimate of the long-term discounted reward $R_i^t$ for each pixel. We train the denoiser for a common noise used in literature, namely additive white Gaussian noise with $\sigma_{denoiser} = 25$. The training process is described in Algorithm~\ref{algo}.

For inference, only the well-trained policy network will be applied.
Taking a noisy image as input, the policy network $\pi$ outputs the probability for selecting an action and the residual to be recovered is sampled greedily with highest probability. 
It should be noted that we add entropy of the policy into the loss, which encourages the agent to explore more possible trajectories. 
We have seen in experiments that the pre-trained DRL denoiser is unable to cope with the artifacts appearing in PnP iterations without adding the entropy as an additional loss term. 
We found that setting the coefficient for the entropy item $\eta$ = 0.01 achieves a promising performance.

\begin{figure*}[!t]
\begin{center}
\begin{tabular}{c}
\includegraphics[width=6.8in]{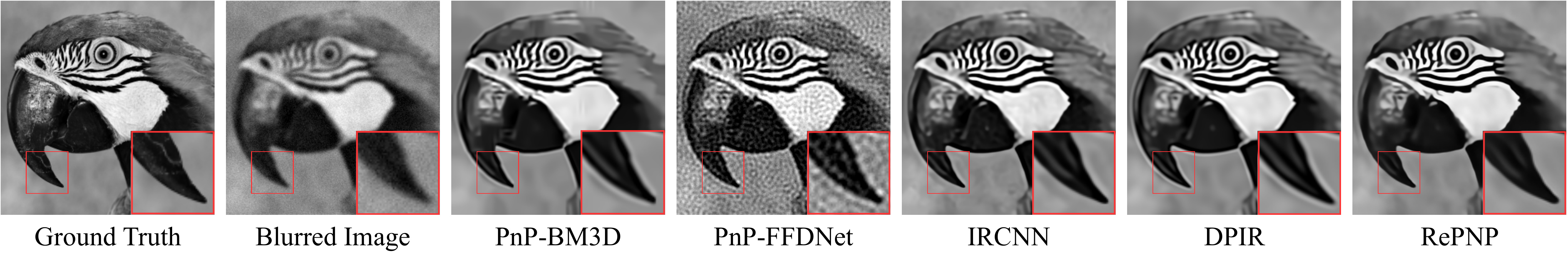}\\
\end{tabular}
\caption{Image deblurring results with a biased choice of forward model in PnP iterations (the standard deviation of the Gaussian kernel for the blurred image is $\sigma_{blur} = 2.0$ and the standard deviation of the Gaussian kernel used in PnP iterations is $\sigma_{est}$ = 2.5).}
\label{deblurimage}
\end{center}
\vspace{-0.2in}
\end{figure*}

\begin{figure*}[!t]
\begin{center}

\begin{tabular}{c}
\includegraphics[width=6.8in]{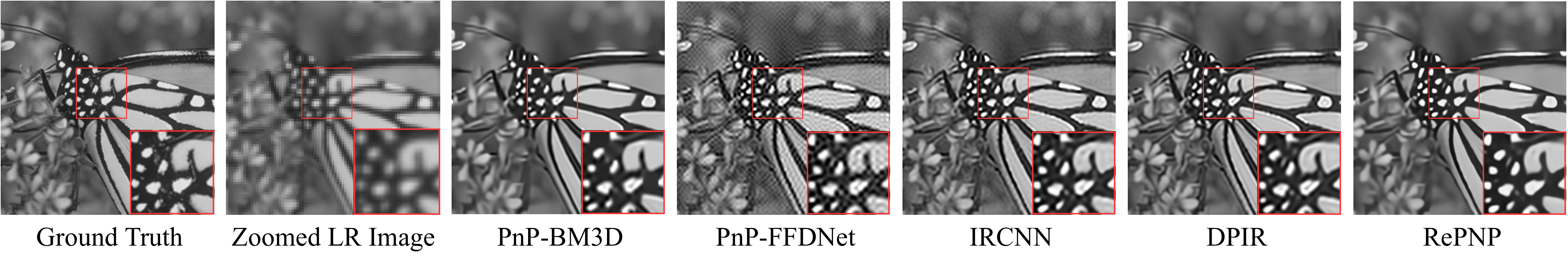}\\

\end{tabular}
\caption{Single image super resolution results with a biased choice of the forward model in PnP iterations (the standard deviation of the Gaussian kernel for the low-resolution image is $\sigma_{LR} = 2.0$, while the standard deviation of the Gaussian kernel for the PnP iteration is $\sigma_{est}$ = 2.3 and the sub-sampling factor is 3).}
\label{srimage}
\end{center}
\vspace{-0.2in}
\end{figure*}

\begin{algorithm}
\hspace*{\algorithmicindent}\noindent \textbf{Input:} initial parameters $\theta_{\pi_0}$ and $\theta_{V_0}$, training epochs $N$,\par
\qquad \qquad \ \; steps $T$
\caption{Training process using PPO with maximum entropy}\label{algo}
\begin{algorithmic}[1]
      \For{n = 0,1,2,$\dots$,N}
        \State Collect set of trajectories $D_n = {J_i^T}$ for each pixel in every
        \Statex \qquad training image by denoising with $\pi_n(\cdot|\theta_{\pi_n})$
        \State Compute rewards-to-go $\hat{R^t}$ for step $t \leq T$
        \State Compute advantage estimates, $\hat{A^t} = \hat{R^t} - V^t$ based on
        \Statex \qquad $V(\cdot|s^t, \theta_{V_n})$
        \State Update the policy by maximizing the PPO-Clip objective:
        \Statex \qquad $\theta_{\pi_{n+1}} = $
         $\mathrm{argmax}_{\theta{\pi}} \frac{1}{|D_n|T}\sum_{J_i \in D_n} \sum_{t=0}^T [ \min(r^t(\theta_\pi)\hat{A^t},$
         \Statex \qquad $\text{clip}(r^t(\theta_\pi),1-\epsilon, 1+\epsilon)\hat{A^t}) + \eta H(\pi(\cdot|s^t, \theta_\pi)) ]$, \Statex \qquad where $r^t = \frac{\pi(a^t|s^t, \theta_\pi)}{\pi(a^t|s^t,   \theta_{\pi_n})}$,  $H(\cdot)$ calculates the entropy,\Statex \qquad and $\text{clip}(\cdot)$ bounds $r^t$ within $[1-\epsilon, 1+\epsilon]$ 
        \State Fit value function by regression on mean-squared error:
        \Statex \qquad $\theta_{V_{n+1}} = \mathrm{argmin}_{\theta_V}\frac{1}{|D_n|T}\sum_{J_i \in D_n} \sum_{t=0}^T (V(v^t|s^t, \theta_{V})$
        \Statex \qquad $-\hat{R^t})^2$
      \EndFor
      \State $\textbf{return} \quad \theta_{\pi_{N+1}}, \  \theta_{V_{N+1}}$
    
\end{algorithmic}
\end{algorithm}

\section{Experiments and Results} \label{section4}
\subsection{Implementation Details}
Our DRL denoiser is trained on 400 images from the Berkeley segmentation dataset (BSD)~\cite{bsd} and we randomly cropped each image into a single patch of size 70 $\times$ 70 during each iteration.
The noisy input is generated by adding additive white Gaussian noise with $\sigma_{denoiser}=25$ to clean patches. 
The hyper parameters for training are set as follows: learning rate of $10^{-4}$; batch size is set to 32, and the termination state is $T=5$. Image rotation and flips are used for data augmentation to reduce overfitting.

For the parameters in PnP, from~\eqref{eq5} and \eqref{eq6}, there are two parameters to tune.
For a certain degradation, $\lambda$ should be kept fixed during the PnP iterations and $\mu$ controls the strength of the denoiser.
Specifically, in \eqref{eq6}, the noise level for the denoiser is correlated to $\sqrt{\lambda/\mu}$ and it should vary from large to small during the PnP iterations. 
In the following experiments, we set $\sqrt{\lambda/\mu}$ decays exponentially from 50 to $\sigma_n$, where $\sigma_n$ is determined by the additive noise level in \eqref{eq1}. 

\vspace{-0.1in}
\subsection{Image Deblurring}\label{sec4.2}
Following the common setting of image deblurring, the blurry images are generated by convolving the original images with a blur kernel and then adding additive white Gaussian noise, as described in~\eqref{eq1}. 
Here, $H$ is a circulant matrix representing the circular convolution with the blur kernel.
We adopt the commonly used isotropic Gaussian kernel of size 25 $\times$ 25 to synthesize blurry images since the blur strength can be directly controlled by the standard deviation of the Gaussian kernel, denoted as $\sigma_{blur}$. In the experiments, the blurry images were generated using blur kernel with $\sigma_{blur}$ = 2.0 and then corrupted with additive white Gaussian noise with standard deviation $\sigma_n$ = 7.65. 

We compare our RePNP with four state-of-the-art image restoration methods, namely PnP-BM3D~\cite{BM3D}, PnP-FFDNet~\cite{ffdnet}, IRCNN~\cite{ircnn} \footnote{IRCNN trains 25 denoisers with different noise levels and implements them according to the increasing $\mu$ during each PnP iteration~\cite{ircnn}.} and DPIR~\cite{dpir}, for image deblurring on Set12~\cite{dncnn}. 
To evaluate the robustness of these methods with respect to drift in the forward model, we implemented a biased choice of the forward model in the PnP iteration \eqref{eq5} by using different blur kernels with $\sigma_{est}$ in the range [2.2, 2.8]. 
The results are measured in terms of average peak signal-to-noise ratio (PSNR) and shown in Table~\ref{tab-deblur}.

The PSNR results in Table~\ref{tab-deblur} show that RePNP outperforms other state-of-the-art denoisers when the blur level is overestimated ($\sigma_{est}$ > $\sigma_{blur}$) in the PnP iterations. 
Besides, as one can see, our RePNP is much more lightweight compared with other conventional deep learning based PnP frameworks. 
From the visual results in Fig.~\ref{deblurimage}, apart from PnP-FFDNet that completely fails to remove the artifacts introduced in the PnP iteration, other methods produce evident artifacts (white outline around the edge) due to the inconsistency of the blur kernel, whereas our RePNP restores a sharper edge without artifacts.

\begin{table}[t]
\caption{Deblurring results for different methods with biased choice of blur kernel in PnP iteration.}
\label{tab-deblur}
\setlength{\tabcolsep}{1.2mm}{
\begin{threeparttable} 
\begin{tabular}{cccccc}
\hline
Method & PnP-BM3D & PnP-FFDNet & IRCNN & DPIR & RePNP \\ \hline
\#Params & - & 0.49M & 4.75M$^*$ & 32.64M & \textbf{0.42M} \\ \hline
$\sigma_{est}$ & \multicolumn{5}{c}{PSNR (dB)} \\ \hline
2.2 & 26.34 & 20.68 & \textbf{26.65} & 26.64 & 26.55 \\ \hline
2.3 & 26.08 & 20.79 & 26.35 & 26.30 & \textbf{26.36} \\ \hline
2.4 & 25.66 & 20.81 & 25.89 & 25.78 & \textbf{25.97} \\ \hline
2.5 & 25.10 & 20.74 & 25.31 & 25.15 & \textbf{25.41} \\ \hline
2.6 & 24.45 & 20.59 & 24.65 & 24.46 & \textbf{24.74} \\ \hline
2.7 & 23.75 & 20.38 & 23.96 & 23.74 & \textbf{24.03} \\ \hline
2.8 & 23.04 & 20.12 & 23.27 & 23.03 & \textbf{23.31} \\ \hline
\end{tabular}
\begin{tablenotes}   
\footnotesize               
\item[*]  The total number of parameters of IRCNN is 0.19M $\times$ 25 = 4.75M.
\end{tablenotes}            
\end{threeparttable}}   
\vspace{-0.2in}
\end{table}

\vspace{-0.1in}
\subsection{Single Image Super Resolution}
The data fidelity term for single image super resolution (SISR) task in \eqref{eq2} can be formulated as $\lVert{SGx-y}\rVert^2_2$, where $S$ and $G$ are the subsampling matrix and a circulant matrix representing the convolution with an anti-aliasing kernel, respectively. 
In this experiment, we use the isotropic Gaussian kernel (same as in Sec.~\ref{sec4.2}) as the anti-aliasing kernel. 
The standard deviation of the Gaussian kernel for the low-resolution image is set to $\sigma_{LR}$ = 2.0.
We generated three different sizes of low resolution images by setting the sub-sampling factor $s$ = \{2, 3, 4\} without additive white Gaussian noise ($\sigma_n=0$). 

The compared methods are the same as in Sec.~\ref{sec4.2} and we implement them for single image super resolution on Set12.
We use a biased choice of $G$ by using different anti-aliasing kernels with $\sigma_{est}$ in the range [2.2, 2.5] in PnP to evaluate the generalization ability of different methods.
\begin{table}[htb]
\caption{Image super resolution results for different methods with biased choice of the anti-aliasing kernel in PnP iterations.}
\label{tab-sr}
\setlength{\tabcolsep}{1.2mm}{
\begin{threeparttable} 
\begin{tabular}{cccccc}
\hline
Method & PnP-BM3D & PnP-FFDNet & IRCNN & DPIR & RePNP \\ \hline
\#Params & - & 0.49M & 4.75M$^*$ & 32.64M & \textbf{0.42M} \\ \hline
$\sigma_{est}$ & \multicolumn{5}{c}{PSNR (dB)} \\ \hline
\multicolumn{6}{c}{Sub-sampling factor = 2} \\ \hline
2.2 & 27.73 & 22.11 & 27.80 & 28.01 & \textbf{28.30} \\ \hline
2.3 & 26.34 & 21.33 & 25.64 & 25.75 & \textbf{26.73} \\ \hline
2.4 & 24.30 & 20.30 & 23.44 & 23.46 & \textbf{24.49} \\ \hline
2.5 & 22.04 & 19.13 & 21.43 & 21.39 & \textbf{22.29} \\ \hline
\multicolumn{6}{c}{Sub-sampling factor = 3} \\ \hline
2.2 & 27.27 & 25.69 & 27.34 & 27.48 & \textbf{27.62} \\ \hline
2.3 & 26.61 & 24.33 & 26.10 & 26.21 & \textbf{26.92} \\ \hline
2.4 & 25.38 & 22.74 & 24.48 & 24.55 & \textbf{25.50} \\ \hline
2.5 & \textbf{23.70} & 21.11 & 22.72 & 22.73 & 23.64 \\ \hline
\multicolumn{6}{c}{Sub-sampling factor = 4} \\ \hline
2.2 & 26.12 & 25.21 & 26.12 & 26.17 & \textbf{26.22} \\ \hline
2.3 & 25.92 & 24.87 & 25.72 & 25.77 & \textbf{26.06} \\ \hline
2.4 & 25.46 & 24.31 & 25.06 & 25.11 & \textbf{25.60} \\ \hline
2.5 & 24.74 & 23.49 & 24.14 & 24.20 & \textbf{24.81} \\ \hline
\end{tabular}
\begin{tablenotes}   
\footnotesize               
\item[*]  The total number of parameters of IRCNN is 0.19M $\times$ 25 = 4.75M.
\end{tablenotes}            
\end{threeparttable}}  
\vspace{-0.2in}
\end{table}
Table~\ref{tab-sr} shows the average PSNR(dB) results for different methods for SISR on Set12. 
It is clear that our RePNP outperforms the competing methods for most of the cases.
From Fig.~\ref{srimage}, our method provides a better visual result without artifacts in the center white region in the red rectangular sub image, while other methods fail to provide a satisfying result. 
Compared with other conventional deep priors, applying DRL denoiser yields more robust results with higher parameter efficiency. 

\section{Conclusions}
In this paper, we proposed a novel deep reinforcement learning based PnP framework, dubbed RePNP, for image restoration.
We considered a more practical yet challenging setting, where the 
observation model is inaccurate.
By modeling the image denoising as a markov decision process, we trained a fully convolutional network via PPO as the image denoising prior in the PnP framework. 
Our RePNP has much less parameters compared with its supervised learning based counterparts and it is also more robust when the observation model used in PnP deviates from the actual one.
Extensive experiments demonstrate that our RePNP can obtain more promising restoration results on image deblurring and super resolution tasks in practice compare with other state-of-the-art image restoration baselines. 

\bibliographystyle{IEEEbib}
\bibliography{DRL_PnP}

\end{document}